\documentclass[prl,reprint]{revtex4-1}
\usepackage{graphicx}
\usepackage{color}
\usepackage{gensymb}
\usepackage{bm}
\usepackage{hyperref}
\usepackage{epstopdf}

\begin{document}

\title{$\nu=1/2$ Fractional Quantum Hall Effect in Tilted Magnetic Fields}
\date{today}

\author{Sukret Hasdemir}
\affiliation{Department of Electrical Engineering,
Princeton University, Princeton, New Jersey 08544}
\author{Yang Liu}
\affiliation{Department of Electrical Engineering,
Princeton University, Princeton, New Jersey 08544}
\author{H.\ Deng}
\affiliation{Department of Electrical Engineering,
Princeton University, Princeton, New Jersey 08544}
\author{M.\ Shayegan}
\affiliation{Department of Electrical Engineering,
Princeton University, Princeton, New Jersey 08544}
\author{L.N.\ Pfeiffer}
\affiliation{Department of Electrical Engineering,
Princeton University, Princeton, New Jersey 08544}
\author{K.W.\ West}
\affiliation{Department of Electrical Engineering,
Princeton University, Princeton, New Jersey 08544}
\author{K.W.\ Baldwin}
\affiliation{Department of Electrical Engineering,
Princeton University, Princeton, New Jersey 08544}
\author{R.\ Winkler}
\affiliation{Department of Physics, Northern Illinois University, DeKalb, Illinois 60115 and\\ Materials Science Division, Argonne National Laboratory, Argonne, Illinois 60439}

\date{\today}

\begin{abstract}
Magnetotransport measurements on two-dimensional electrons confined to wide GaAs quantum wells reveal a remarkable evolution of the ground state at filling factor $\nu=1/2$ as we tilt the sample in the magnetic field. Starting with a compressible state at zero tilt angle, a strong $\nu=1/2$ fractional quantum Hall state appears at intermediate angles. At higher angles an insulating phase surrounds this state and eventually engulfs it at the highest angles. This evolution occurs because the parallel component of the field renders the charge distribution increasingly bilayer-like.
\end{abstract}

\maketitle

The fractional quantum Hall effect (FQHE) is observed in clean two-dimensional electron systems (2DESs) under high perpendicular magnetic fields ($B_{\perp}$) \cite{TsuiPRL1982}, usually at odd-denominator Landau level (LL) filling factors $\nu$ \cite{JainCF2007}. In the excited LL ($N=1$), there is also a FQHE state at the even-denominator filling $\nu=5/2$ \cite{WillettPRL1987}; this state is likely a Moore-Read (Pfaffian) \cite{MooreNPB1991} state and has potential use for topological quantum computing thanks to its non-Abelian statistics \cite{NayakRMP2008}.
Even-denominator FQHE is also seen in the ground-state ($N=0$) LL in systems with bilayer-like charge distributions at $\nu=1/2$ \cite{SuenPRL1992a, EisensteinPRL1992, SuenPRL1992b, SuenThesis, SuenPRL1994, ManoharanPRL1996, ShayeganSST1996, LayPRB1997, ShayeganTALDS, LuhmanPRL2008, ShabaniPRL2009a,ShabaniPRL2009b,ShabaniPRB2013}, $\nu=3/2$ \cite{SuenPRL1994}, and $\nu=1/4$ \cite{LuhmanPRL2008, ShabaniPRL2009a}. The $1/2$ FQHE is observed in either double GaAs electron quantum wells (QWs) \cite{EisensteinPRL1992} or in wide GaAs QWs \cite{ SuenPRL1992a, SuenPRL1992b, SuenThesis, SuenPRL1994, ManoharanPRL1996, ShayeganSST1996, LayPRB1997, ShayeganTALDS, LuhmanPRL2008, ShabaniPRL2009a,ShabaniPRB2013,ShabaniPRL2009b} where the repulsion between the electrons makes the charge distribution bilayer-like; it has also been reported very recently in 2D hole systems confined to relatively wide GaAs QWs \cite{LiuPRB2014,LiuPRL2014}, and in bilayer graphene \cite{KiNanoLett2014}. Although the $\nu=1/2$ FQHE in wide QWs might also be a Pfaffian state \cite{GreiterPRL1991}, it is more likely described by the Abelian, two-component, Halperin-Laughlin $\Psi_{331}$ wavefunction where the components are the two layers, or equivalently, the symmetric and antisymmetric electric subbands \cite{Y8, Y9, Y11, Y13, Y14, Y15, Y16,ScarolaPRB2010,ThiebautPRB2014}. This state is stable when the interlayer tunneling, quantified by the symmetric-antisymmetric subband energy separation $\Delta_{\text{SAS}}$, is much smaller than the intralayer Coulomb interaction energy, and the interlayer and intralayer Coulomb interaction energies are comparable. 

In 2DESs confined to wide QWs at $\nu=1/2$, an evolution from a compressible to a FQHE, and finally to an insulating phase (IP) is observed as the density ($n$) is increased \cite{ManoharanPRL1996}. Increasing $n$ causes $\Delta_{\text{SAS}}$ to decrease while the intralayer Coulomb energy increases, allowing the two-component $\nu=1/2$ FQHE to be stabilized, but this state is eventually destroyed by an IP when the system splits into two weakly-coupled layers at very high $n$. The role of a parallel magnetic field ($B_\|$), however, has not been systematically studied. An early study \cite{SuenPRL1992a} indicated that the application of $B_\|$ causes the FQHE to get destroyed by an IP, but it has also been reported that $B_\|$ can strengthen a weak $\nu=1/2$ FQHE \cite{LayPRB1997, LuhmanPRL2008}. 

Here we report, for 2DESs confined to a wide GaAs QW, the full evolution of the ground state at $\nu=1/2$ through compressible, FQHE, and IP as we increase $B_\|$ while keeping $n$ constant.
This is similar to the evolution seen in a wide QW at $B_\|=0$ where raising $n$ makes the system increasingly bilayer-like with reduced interlayer tunneling. We explain the similarity by showing that the large $B_\|$ also makes the 2DES bilayer-like and lowers the interlayer tunneling. Moreover, the $\nu=1/2$ FQHE we observe at high $B_\|$ can appear at much smaller interlayer distances compared to the $B_\|=0$ case because a large $B_\|$ leads to a smaller effective layer-thickness which results in a stronger intralayer Coulomb interaction.

We studied 2DESs confined to 60- and 65-nm-wide GaAs QWs grown by molecular beam epitaxy. The QWs are flanked by undoped AlGaAs spacer layers and Si $\delta$-doped layers. The samples were 4 $\times$ 4 mm$^2$, with In:Sn contacts at four corners, and each was fitted with a Ti/Au front-gate and an In back-gate, allowing us to make the charge distribution symmetric \cite{SuenPRL1994, SuenThesis, ManoharanPRL1996, ShayeganTALDS, ShabaniPRL2009a, LiuPRB2011,ShayeganSST1996,LayPRB1997} and also vary the 2DES density $n$, which we give throughout this report in units of $10^{11}$ cm$^{-2}$. We studied three samples with as-grown densities $0.45$, $1.4$, and $2.4$. We made measurements in a dilution refrigerator with a base temperature $T \approx$ 30 mK, an 18 T superconducting magnet, and a tilting stage so that the sample normal could be tilted at an angle ($\theta$) with respect to the magnetic field.

Figure 1 shows a series of longitudinal ($R_{xx}$) and Hall ($R_{xy}$) resistance traces obtained at different $\theta$ for $n=1.4$ in a 65-nm-wide QW. At $\theta=0$ there is no $\nu=1/2$ FQHE but at $\theta=10\degree$ a $\nu=1/2$ FQHE appears and becomes stronger as we further increase $\theta$. The $\nu=1/2$ FQHE is also signaled by the quantization of $R_{xy}$. At higher $\theta$ an IP appears at low $\nu$ and moves to higher $\nu$ with increasing $\theta$. At $\theta=35\degree$ the IP starts just to the right of the $\nu=1/2$ FQHE and, at $\theta=37\degree$, reentrant IPs flank a strong $\nu=1/2$ FQHE on both sides. At yet higher $\theta$, the entire $\nu=1/2$ region is covered by the IP. In Fig. 1 inset we show the charge distributions calculated at $B_{\perp}=0$ at $n=1.4$ for $B_\|=0$, $5.4$, $9.7$ T, corresponding to $\theta=0, 25\degree, 40\degree$ at $\nu=1/2$, illustrating the increasing bilayerness of the charge distribution with tilt.

\begin{figure}
\includegraphics[width=0.42\textwidth]{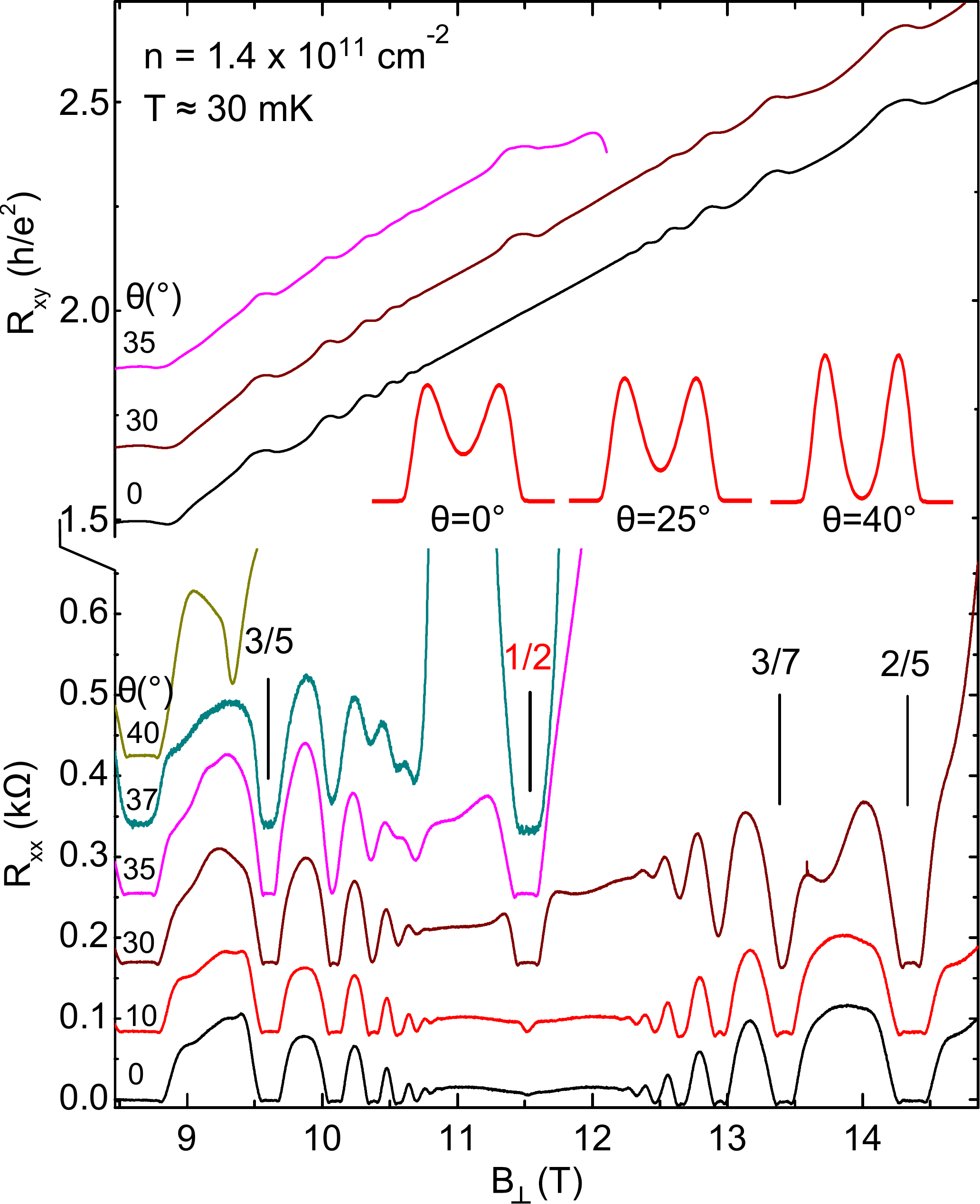}
\caption{$R_{xx}$ and $R_{xy}$ traces around $\nu=1/2$ for electrons confined to a 65-nm-wide QW at $n=1.4$ are plotted as a function of $B_{\perp}$ for several tilt angles. Traces are shifted vertically for clarity. Inset: Charge distributions calculated self-consistently at the indicated angles, but taking into account only the parallel component of the magnetic field.}
\end{figure} 

Figure 2 shows data for three different samples. The traces in Fig. 2(a), taken at a high density of 2.0 in a 65-nm-QW, show a strong $\nu=1/2$ FQHE at $\theta=0$, consistent with previous, density-dependent studies \cite{ShabaniPRB2013}. Tilting the sample causes the $\nu=1/2$ FQHE to disappear and turn into an IP for $\theta > 20\degree$. At $n=0.86$ (Fig. 2(b)), qualitatively similar to Fig. 1 data, the ground state at $\nu=1/2$ is compressible at $\theta=0$, turns into a FQHE in a relatively small range of $\theta$ near $50\degree$, and then becomes insulating at larger $\theta$. In the lowest density data, taken at $n=0.45$ in a 60-nm-QW (Fig. 2(c)), a developing $\nu=1/2$ FQHE is seen only in a very small range of $\theta$ near $70\degree$, and is replaced by an IP at higher $\theta$.

\begin{figure*}
\includegraphics[width=0.8\textwidth]{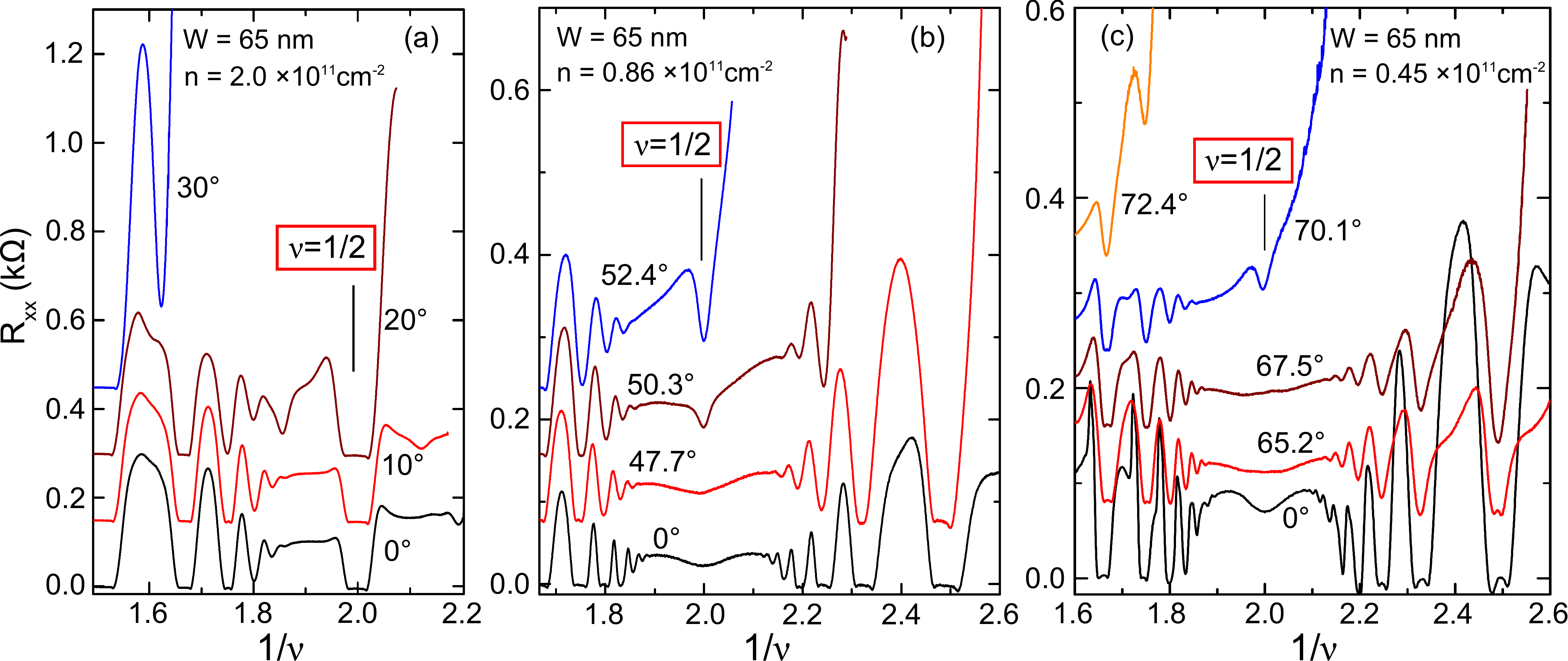}
\caption{$R_{xx}$ vs $1/\nu$ traces around filling factor $1/2$ for electrons confined to 65- and 60-nm-wide GaAs QWs at densities $2.0$, $0.86$ and $0.45$ for several tilting angles. In each panel traces are shifted vertically for clarity.}
\end{figure*}

The evolution we observe as a function of increasing $\theta$ is qualitatively similar to the one seen at $\theta=0$ in a wide QW with a fixed width when $n$ is increased \cite{SuenPRL1992b,SuenThesis,SuenPRL1994,ManoharanPRL1996, ShabaniPRB2013,ShayeganSST1996, ShayeganTALDS}. The samples we study here in fact provide an example of such behavior. The bottom ($\theta=0$) traces in Figs. 2(a) and 2(b) show that the ground state at $\nu=1/2$ is compressible at low density but turns into a strong FQHE state as we increase $n$. The $\theta=0$ data at higher $n$ (not shown here) reveal that, in a very narrow range of $n$, there is a FQHE at $\nu=1/2$ which is flanked on \textit{both} sides, i.e., slightly higher and lower $B_{\perp}$, by IPs \cite{ManoharanPRL1996}. Such a reentrant behavior is also seen in our tilted field data in Fig. 1 where the $R_{xx}$ trace at $\theta=37\degree$ shows IPs on both sides of a strong $\nu=1/2$ FQHE. The IPs surrounding the $\nu=1/2$ FQHE are likely to have interlayer correlations and are interpreted as pinned, bilayer Wigner crystal states \cite{ManoharanPRL1996,ShayeganSST1996,ShayeganTALDS}. 

We made measurements in the 65-nm-wide QWs at numerous densities and summarize our data in Fig. 3 in a simple $B_\|$ vs density "phase diagram." At $B_\|=0$, the ground state is compressible for $n<1.4$, FQHE for $1.4<n<2.1$, and insulating for $n$ slightly higher than $2.1$; this is consistent with density-dependent studies \cite{SuenPRL1994,ManoharanPRL1996,ShabaniPRB2013,ShayeganTALDS}. Tilting the sample causes the compressible states to turn into FQHE, and this happens at a higher $B_\|$ for lower $n$. If we tilt the sample further, the $\nu=1/2$ FQHE is eventually destroyed by an IP. 

\begin{figure}
\includegraphics[width=0.36\textwidth]{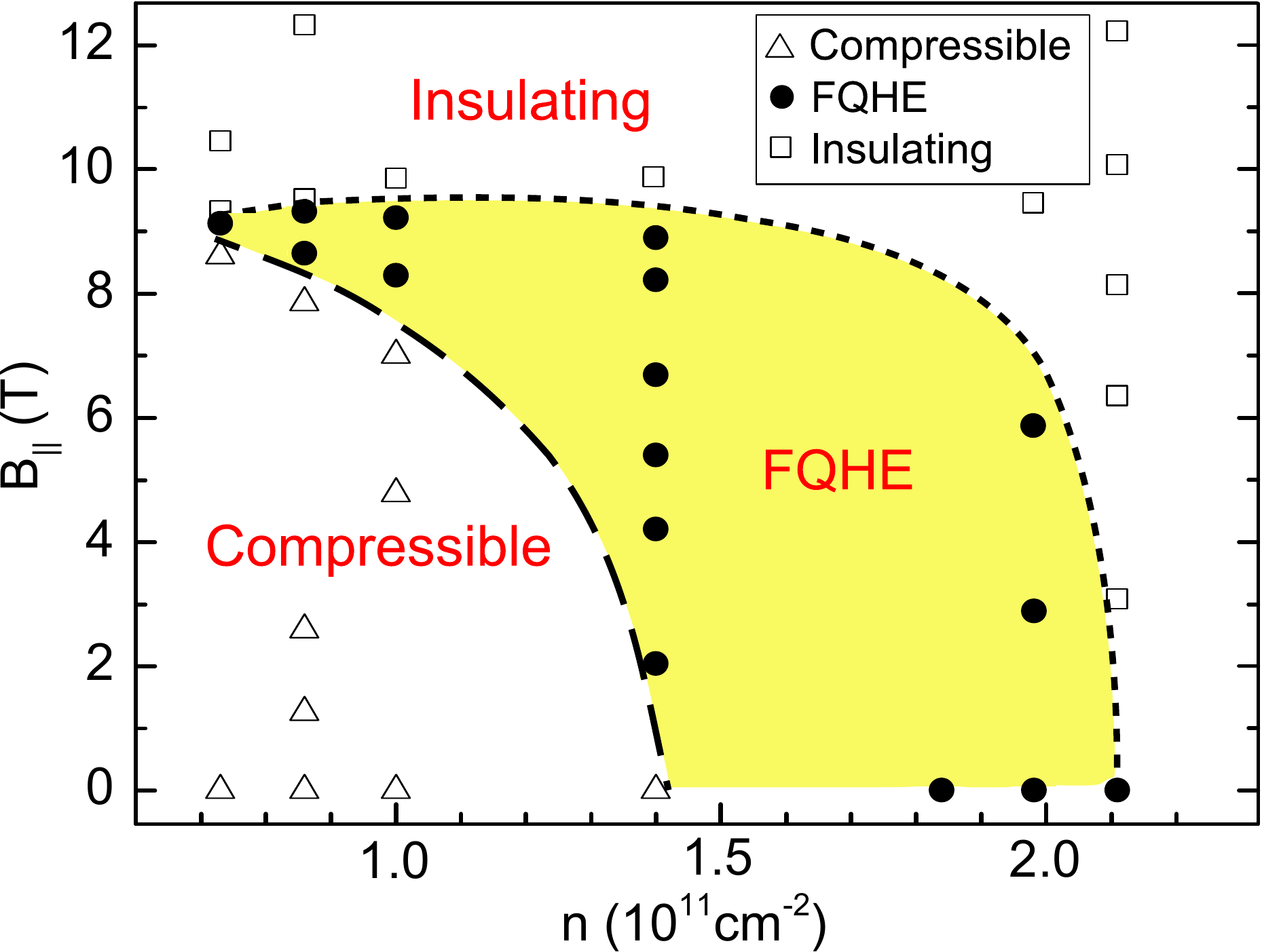}
\caption{A $B_\|$ vs density phase diagram for the ground states at $\nu=1/2$ based on data from 65-nm-wide QWs.}
\end{figure}

In order to understand our tilted field data and also to make a quantitative comparison to the $B_\|=0$ results, we consider two parameters that are often used to characterize the 2DESs in a wide QW at $\nu=1/2$: $d/l_B$ and $\alpha=\Delta_{\text{SAS}}/(e^2/4 \pi \epsilon l_B)$\cite{SuenPRL1994, ManoharanPRL1996, ShayeganSST1996, LayPRB1997, ShayeganTALDS, LuhmanPRL2008, ShabaniPRL2009a,ShabaniPRB2013, LiuPRB2014,LiuPRL2014,Y11, Y13, Y14, Y15, Y16,SuenThesis,ShabaniPRL2009b}. The ratio $d/l_B$, where $d$ is the distance between the two peaks of the charge distribution (i.e., the ``interlayer distance'') and $l_B$ is the magnetic length for $B_{\perp}$, is the ratio of the intralayer and interlayer Coulomb energies, $e{^2}/ 4 \pi \epsilon l_{B}$ and $e^{2}/4 \pi \epsilon d$. The parameter $\alpha$ is the ratio of the interlayer tunneling energy, quantified in $B_\|=0$ studies by the symmetric-antisymmetric subband gap $\Delta_{\text{SAS}}$, to the intralayer Coulomb energy. In density-dependent studies of wide GaAs QWs, the $\nu=1/2$ FQHE is observed when $0.05 \lesssim \alpha \lesssim 0.15$ and $5 \lesssim d/l_B \lesssim 8$ \cite{ShabaniPRB2013}. 

To compare our results to such studies, we performed self-consistent calculations of the charge distributions and subband dispersions for a 65-nm-wide GaAs QW at $B_{\perp}=0$, both at zero and finite values of $B_\|$ in a large range of densities. The calculations at $B_\|=0$ directly provide tunneling energies  ($\Delta_{\text{SAS}}$) at a given density. For $B_\|>0$, however, simply taking the difference between the energies of the lowest two electric subbands does not give the interlayer tunneling because the vector potential $A(z)$ due to $B_\|$ acts as an additional $k_\|$-dependent confining potential so that a moderate $B_\|$ of a few Tesla completely depopulates the first excited electric subband.  The bilayer-like charge distribution in real-space then corresponds to two local minima in reciprocal-space of the energy dispersion for the lowest electric subband \cite{Jungwirth1993}. To estimate the relevant, effective tunneling energy, we use the charge distribution calculations in the following manner \cite{Note1,Note2}. We first define a parameter $\beta$ as the ratio of the height at the midpoint between the two peaks in the charge distribution to the height at the peaks; $\beta$ is closely related to the interlayer tunneling. Next, we make a plot of $\beta$ vs $\Delta_{\text{SAS}}$ for the $B_\|=0$ calculations in a large density range, and then use this plot to estimate the effective $\Delta_{\text{SAS}}$ for the charge distributions calculated at $B_\|>0$ by using their $\beta$ values. In this procedure, we are effectively matching the charge distribution calculated at a given density and $B_\|$ to one calculated at $B_\|=0$ but at a higher density. 

In Fig. 4 top panel we present a phase diagram for our data from the 65-nm-wide QWs with axes $\alpha=\Delta_{\text{SAS}}/(e^2/ 4 \pi \epsilon l_B)$ and $d/l_B$, where $\Delta_{\text{SAS}}$ used for the $B_\|\ne0$ data points are based on the procedure described in the preceding paragraph. The diagram shows the parameter ranges where we observe the three phases of the 2DES at $\nu=1/2$: compressible, FQHE, and insulating. To illustrate some details of the diagram, we have highlighted three points, $A$, $B$, and $C$, for which we show the calculated charge distributions in the lower panels of Fig. 4. Case $A$ is at low density ($n=1.0$) with $B_\|=0$, has an essentially single-layer-like charge distribution, and is compressible at $\nu=1/2$. Case $B$ is at higher density ($n=1.8$), also with $B_\|=0$, has a more bilayer-like charge distribution, and exhibits a $\nu=1/2$ FQHE. The thin solid curve in the phase diagram going through the points $A$ and $B$ indicates how the 2DES evolves as $n$ is increased at $B_\|=0$. Case $C$ is at low density ($n=1.0$) but has $B_\|=8\, \text{T}$, and shows a FQHE at $\nu=1/2$. The curve connecting $A$ and $C$ goes through data points all taken at $n=1.0$, and illustrates how the ground state evolves from compressible to FQHE to insulating as $B_\|$ is increased. Note that as we go from $A$ to $B$, or $A$ to $C$, the system becomes increasingly bilayer-like as density, or $B_\|$, is increased. 

In the phase diagram of Fig. 4 we mark, in blue, the region where the $\nu=1/2$ FQHE is observed in density-dependent studies which were conducted at $B_\|=0$ \cite{ShabaniPRB2013}. It is clear that the $\nu=1/2$ FQHE is observed at significantly smaller values of $d/l_B$ and $\alpha$ when $B_\|>0$ compared to the $B_\|=0$ case. 

The difference in $d/l_B$ can be explained based on the softening of the intralayer Coulomb interaction because of the non-zero electron layer thickness $\lambda$ which we define as the full-width-at-half-maximum of the charge distribution for each ``layer'' (see the lower left panel of Fig. 4). Note that the $\Psi_{331}$ state is theoretically expected to be stable when the intralayer and interlayer Coulomb energies are comparable \cite{Y9,Y11}. For an ideal bilayer system (with zero layer thickness), the ratio $d/l_B$ accurately reflects the relative strengths of the intralayer and interlayer Coulomb interactions and the $\Psi_{331}$ FQHE at $\nu=1/2$ should be observable for $d/l_B\sim 2$ (note that $l_B$ is the magnetic length set by $B_{\perp}$). However, when $\lambda$ is comparable to or larger than $l_B$, the short-range component of the Coulomb interaction, which is responsible for the FQHE, softens \cite{ShayeganPRL1990, HePRB1990}. Associating the $\nu=1/2$ FQHE with the $\Psi_{331}$ state, it is thus not surprising that we see the FQHE in the presence of a high $B_\|$ at a smaller $d/l_B$ ($\simeq 3.5$) compared to the $B_\|=0$ case ($d/l_B>5$, see Fig. 4 phase diagram): $B_\|$ introduces additional confinement, so that for a given $n$, $\lambda/l_B$ is smaller at finite $B_\|$. The short-range component of the intralayer interaction is stronger when $B_\|$ is large (e.g., $\lambda/l_B\simeq 1.8$ for case $C$ in Fig. 4) compared to the $B_\|=0$ situation (e.g., $\lambda/l_B\simeq 2.7$ for case $B$ in Fig. 4); therefore to ensure the proper intralayer to interlayer interaction ratio which favors the $\Psi_{331}$ state, a relatively stronger interlayer interaction (larger $e^2/ 4 \pi \epsilon d$) is also needed, implying a smaller $d/l_B$
\cite{SuenPRL1994, ShabaniPRB2013}. In a sense, because of its smaller layer thickness, the 2DES in a wide GaAs QW at large $B_\|$ is closer to an ideal bilayer system. This is similar to the case of 2D holes (at $B_{\|}=0$) where the larger effective mass leads to a smaller layer thickness and the $\nu=1/2$ FQHE is observed for $d/l_B=3.6$ and $\lambda/l_B=1.4$ \cite{LiuPRL2014}.

\begin{figure}
\includegraphics[width=0.36\textwidth]{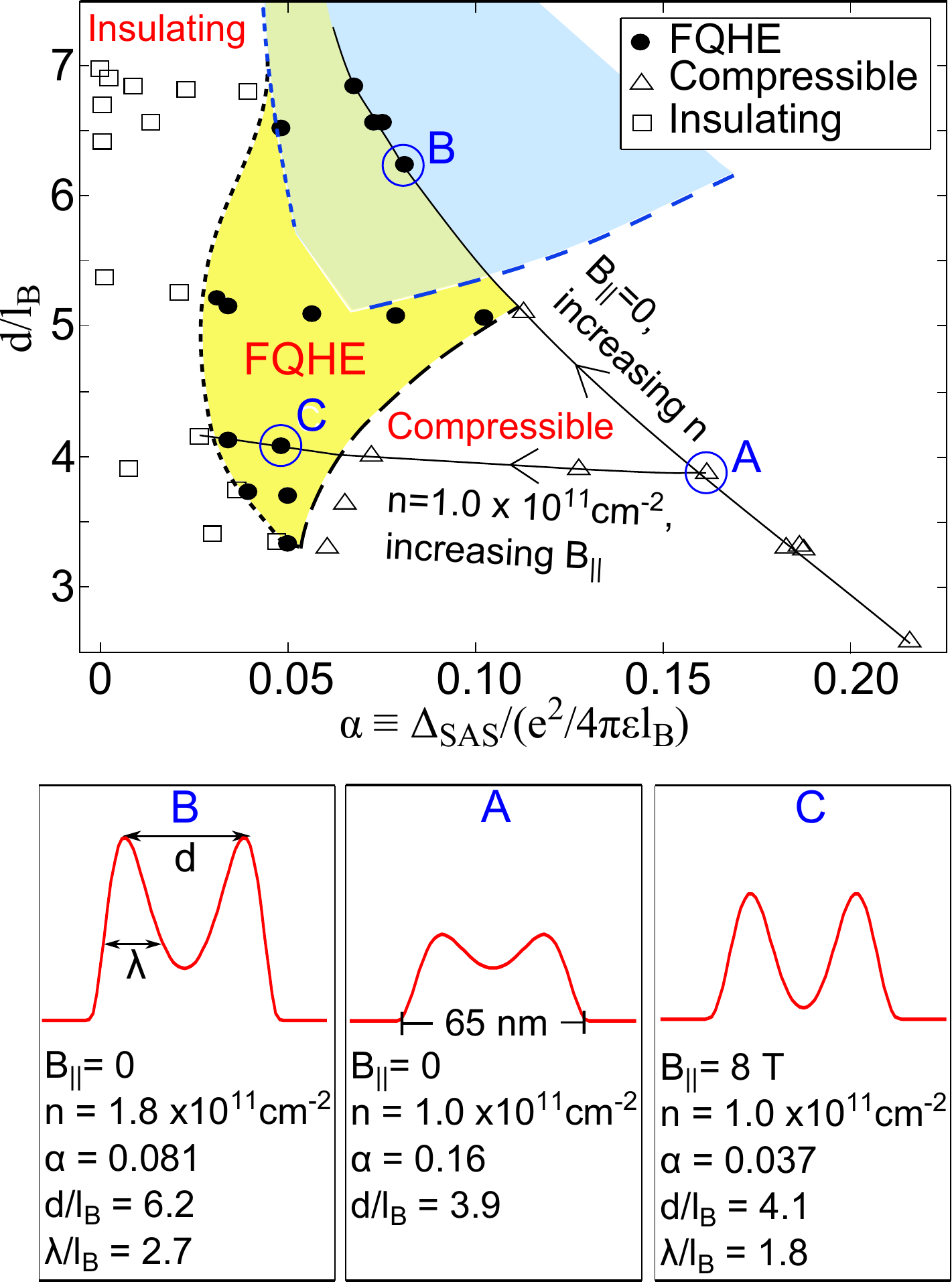}
\caption{Top panel: A $d/l_B$ vs $\alpha$ phase diagram for the ground states at $\nu=1/2$ in a 65-nm-wide QW. The blue area shows the region where the $\nu=1/2$ FQHE is observed in previous studies where $B_\|=0$ and the parameters $\alpha$ and $d/l_B$ were tuned by changing the density in samples with different QW widths \cite{ShabaniPRB2013}. Lower panels: Self-consistently calculated charge distributions for the three cases marked $A$, $B$, and $C$ in the phase diagram. }
\end{figure} 

Finally, we discuss another major difference between the evolutions of the $\nu=1/2$ state seen as a function of $n$ or $\theta$. In $B_\|=0$ experiments, the ground state at $\nu=1/2$ is compressible at very high densities, consistent with two uncorrelated layers each at $\nu=1/4$ \cite{ManoharanPRL1996}. In contrast, in $B_\|\ne0$ experiments at large values of $\theta$, an IP dominates the $\nu=1/2$ region and even extends to higher fillings near 2/3, i.e., near 1/3 filling of each layer. One possible explanation is that, because $d/l_B$ is smaller at the highest $B_\|$ (compared to the highest $n$), interlayer interactions are preserved and the ground state is a pinned, bilayer Wigner crystal. We note that, in single-layer 2DESs confined to very narrow GaAs QWs, IPs which are reentrant around $\nu=1/3$ instead of $\nu=1/5$ have been reported and interpreted as possible Wigner crystal states \cite{YangPRB2003} although the role of disorder and interface roughness is unclear \cite{LuhmanPhysicaE2008}. In our experiments, too, we cannot rule out the possibility that disorder is enhanced at high $B_\|$ because of the increase in the magnitude of the total magnetic field, or because the electron layers are pushed closer to the walls of the QW, similar to the case of 2DESs in narrow QWs.

\begin{acknowledgments}
We acknowledge support through the NSF (DMR-1305691, DMR-1310199 and MRSEC DMR-0819860),  the DOE BES (DE-FG02-00-ER45841), the Gordon and Betty Moore Foundation (Grant GBMF4420), and the Keck Foundation. Work at Argonne was supported by DOE BES (DE-AC02-06CH11357). Our work was partly performed at the National High Magnetic Field Laboratory, which is supported by NSF (DMR-1157490), the State of Florida, and the DOE. We thank S.~Hannahs, T.~P.\ Murphy, G.~E.\ Jones, E.~Palm, J.~H.\ Park, and A.~Suslov for assistance.
\end{acknowledgments}

\bibliographystyle{apsrev4-1}

\end{document}